# Imaging of optically active defects with nanometer resolution

Jiandong Feng[1], Hendrik Deschout[1], Sabina Caneva[2], Stephan Hofmann[2], Ivor Lončarić[3], Predrag Lazić[3] and Aleksandra Radenovic[1]

[1]Laboratory of Nanoscale Biology, Institute of Bioengineering, School of Engineering, EPFL, 1015 Lausanne, Switzerland
[2]Department of Engineering, University of Cambridge, JJ Thomson Avenue, CB3 0FA Cambridge, United Kingdom
[3]Institut Ruđer Bošković, Bijenička 54, 10000 Zagreb, Croatia

*Correspondence should be addressed jiandong.feng@epfl.ch and aleksandra.radenovic@epfl.ch

**Abstract**

Point defects significantly influence the optical and electrical properties of solid-state materials due to their interactions with charge carriers, which reduce the band-to-band optical transition energy. There has been a demand for developing direct optical imaging methods that would allow *in-situ* characterization of individual defects with nanometer resolution. Here, we demonstrate the localization and quantitative counting of individual optically active defects in monolayer hexagonal boron nitride using single molecule localization microscopy. By exploiting the blinking behavior of defect emitters to temporally isolate multiple emitters within one diffraction limited region, we could resolve two defect emitters with a point-to-point distance down to ten nanometers. The results and conclusion presented in this work add unprecedented dimensions towards future applications of defects in quantum information processing and biological imaging.

.

# INTRODUCTION

Vacancy defects are of crucial importance in determining both transport and optical properties of semiconductors. Through interaction with excitons, defects can be optically active at energies lower than the band-to-band optical transition energy, as revealed in their photoluminescence experiments(*1*). Defects in diamond(*2*) and two-dimensional materials(*3*) have been demonstrated to be able to serve as single photon emitters, which are essential for quantum-information processing(*4*). Due to their extraordinary quantum mechanical properties in sensing the local electromagnetic field and temperature, defects are also considered as promising candidates for biological imaging at the nanoscale(*5*). Mapping and localizing individual defects at high spatial resolution will, therefore, add unprecedented dimensions to both applications.

Scanning tunneling microscope (STM) has long been used to directly visualize single defects and their electronic structures. With STM, manipulation of individual point defects can be realized via tip-bias-induced electrochemical reactions(*6*). Advance in aberration-corrected transmission electron microscope (TEM) greatly improved the resolution in characterizing individual defects(*7, 8*). Nevertheless, both methods require special sample preparations and are still not optimal for fast *in-situ* operations. Therefore, development of direct optical technology that images individual defects would match the emerging requirements in both quantum information processing and bioimaging. However, the optical detection and control are hampered by the diffraction limit, which can accumulate multiple defect emitters for defective materials. Super-resolution microscopy techniques can in principle address these problems. Stimulated emission depletion (STED) microscopy has demonstrated a capability of imaging nitrogen–vacancy (N-V) centers in nanodiamond with a resolution of a few nanometers(*9, 10*). Single-molecule localization microscopy (SMLM), based on localizing sparse sets of switchable fluorescent molecules(*11, 12*), has also been used for wide-field parallel imaging of N-V centers in diamond(*13, 14*).

Recently, 2D materials have been shown to host defects that serve as single photon emitters at cryogenic temperatures for transition metal dichalcogenides(*15-18*) and room temperature for hexagonal boron nitride (h-BN)(*19*). In quantum-information processing, optically active defects in 2D materials are attractive since inherently they don't require total internal reflection. In addition, optically active defects have high light extraction efficiency, and electrical connections can be easily integrated(*3*). Therefore, imaging them with super-resolution techniques is of high significance for the realization of reading large-scale networks. The operation at room temperature, together with the high brightness and contrast and defective nature of h-BN makes it the most suitable 2D candidate for room-temperature quantum computing(*3*).

Here we demonstrate the localization and spatial mapping of individual defect emitters in monolayer h-BN by SMLM. This approach allows us to resolve the most precisely localized defect emitters at separations around 10 nm.

Monolayer h-BN comprises alternate boron and nitrogen atoms in a two-dimensional honeycomb arrangement. In order to clarify the structure of atomic defects in h-BN, we first imaged the h-BN samples by aberration-corrected TEM, as shown in **Fig. 1A.** The uniform orientation of all observed defects excludes the possibility of coexistence between boron and nitrogen monovacancies due to the threefold symmetry of h-BN(*8*). Boron monovacancies, further confirmed by the line profile of the image contrast (**Fig. 1B**), is found to be dominant for the h-BN materials(*8*). Because of the electron knock-on effect during TEM irradiation coupled with the beam-induced etching with residual water or oxygen present in the system, more defects can be directly introduced and enlarged to triangular pores(*20*), as shown in **Fig. 1C**.

Photoluminescence experiments on h-BN specimens that were not subjected to electron radiation TEM suggest that the boron monovacancies are intrinsic (**Fig. 1D**), which may form during the growth but as well during the electrochemical transfer process(*21*). Two kinds of emissions are observed in our experiments: the weak homogeneous fluorescence that follows the entire triangular shaped h-BN monolayer(*21*) with an emission peak centered at 580 nm and the ultra-bright individual emitters with emission peak centered at 620 nm (**Fig. 1F**). As defect-free monolayer h-BN has a large bandgap (around 5-6 eV), we speculate that two candidates for the observed emission could be either boron or nitrogen monovacancies. Due to the uncertainty in calculating their electronic structure using the density functional calculations(*22, 23*), both are reasonable candidates for the observed spectrum. Possible causes for this uncertainty are discussed in supporting information. However, according to the TEM results (**Fig. 1a**) and considering the relative density, in monolayer h-BN we have detected **mostly** boron vacancies that may be responsible for an observed optical activity. We infer that the observed weak homogeneous emission is due to the high density of boron monovacancies with neutral charges $V_B^0$ in h-BN, and the ultra-bright emitters are negatively charged boron monovacancies $V_B^-$, as proposed in **Fig. 1E**. As shown in the emission spectrum **Fig.1F**, fluorescence emission above 600 nm is mainly coming from $V_B^-$. We could not avoid the averaging effect in our photoluminescence experiments due to difficulties in isolating individual defects. Better spectroscopic results could be achieved with isolated individual emitters under low-temperature conditions(*23*). Other candidates for the observed defect fluorescence can be boron-nitrogen vacancy(*19*) or three boron plus one nitrogen vacancy. The unique and uniform defect structure of h-BN (**Fig. 1a**) is crucial for the scaling up of a parallel quantum sensor array via selective defect patterning.

Photoswitching was observed after exposure to green light illumination (561 nm). One of the possible explanations for this switching behavior can refer to photo-induced ionization and recombination of $V_B$ between its neutral state and negative state(*24*). We then verified the wavelength dependence (405 nm, 488 nm, and 561 nm) of defect charge state dynamics and found maximum switching rates at 561 nm. No switching event was observed at 405 nm. The charge conversion process can be understood as $V_B^0$ / $V_B^{-1}$ is first excited and then an electron is captured from conduction band for ionization ($V_B^0$ to $V_B^{-1}$) or from valence band for recombination ($V_B^{-1}$ to $V_B^0$), consistent with the evidence reported for both optical excitation of defect transitions in h-BN(*25*) and STM measurements on the excited h-BN defects(*26*).

This photoswitching behavior sets the playground for optical super-resolution reconstruction, which is based on the temporal isolation of densely packed multiple emitters within one diffraction limited region. SMLM trades temporal resolution for spatial resolution, and therefore it requires an ideal imaging sample to be spatially fixed and temporally static(*11*). Both requirements are fulfilled by the solid-state defective h-BN samples. In addition, in this case, no labeling is involved, and the strong emitters are extremely bright and do not bleach. Direct imaging of the sample itself makes it an ideal test case for achieving the fundamental limit of SMLM resolutions(*11*).

Initial image frames typically consisted of dense weak fluorescence background presumably dominated by the much larger population of boron monovacancies still in the inactivated state $V_B^0$. After initial activation, a lot of asynchronous blinking events between two states were observed. Due to the brightness of solid-state emitters, we achieved a high localization precision (*11*) from strong emitters. For example, a localization precision of 3.5 nm is obtained for the emitters shown in **Fig. 2C**. Giving an average localization precision is challenging due to the merged broad distribution of weak and bright emitters. Reconstruction by summing up all temporally isolated Gaussian location distributions of individual emitters defined by their coordinates and localization precision yields **Fig. 2 A and B**. In contrast, a diffraction limited image is produced by summing all image stacks to overlap their signals (**SI Fig. 2**).

The ability to distinguish two closest emitters also imposes stringent requirements on stage drift correction. Precise drift correction can be achieved with an active feedback-controlled system(*27, 28*) or post-imaging processing method using fiducial markers(*11*). We used a large number of fiducial markers to achieve an accurate drift correction. In a representative experiment, trajectories of 27 fiducial markers are used to calculate the averaged drift trajectory and remaining drift error after correction. To demonstrate the imaging capability of solid-state SMLM for individual defects, we first employed Fourier ring correlation to estimate the resolution(*29*), and we obtained 46 nm (**SI Fig. 3**). However, in this case, individual emitters, as well as their distinguishability are more relevant instead of revealing the structure itself. As shown in **Fig. 2C**, high localization precision, and accurate drift correction allowed distinguishing two neighboring emitters with a separation of 10.7 nm. We would like to note that the emitters exhibit a broad distribution of their photon counts (**SI Fig. 4**), thus this resolving capability is dictated by the photophysics and density of strong emitters. Direct optical imaging also allows for *in-situ* characterization of defects under ambient conditions, for example, probing their chemistry and dynamics in different pH environments (**SI Fig.5**), surface depositions (**SI Fig. 6**) and isotopic solvents (**SI Fig.7**). Among these conditions, the fluorescence signals were found to be sensitive to acidity of the environment.

Compared to TEM and STM, for *in-situ* operation under ambient conditions, SMLM introduces minimum destruction to the sample. The same argument also holds for the high power density used for the depletion laser beam in STED(*9*). In addition, standard SMLM has a relatively large field of view, adequate for the imaging of single-crystal 2D material grains. Recently, novel SMLM modalities (*30, 31*) have extended the field of view considerably. The spatial resolution of our results can still be improved using better drift correction methods(*28*) while a better temporal resolution can be achieved by a spatio-temporal cumulant analysis of the image sequence in super-resolution optical fluctuation imaging (SOFI)(*32*), shown in **SI Fig. 8.** SOFI would be particularly interesting for revealing the dynamical processes on defects at the relevant time scales.

Although we are imaging the structure itself without labeling, solid-state SMLM can also provide additional information for revealing chemical contrast of defect types owning to the imaging capability with multiple color channels, as shown in the composited **Fig. 3A**. The dual color imaging led to a preferred spatial distribution of strong emitters near the edge under red channel (**Fig. 3B** and **D**, laser 561 nm) while no spatial preference was found for green channel (**Fig. 3C** and **E**, laser 488 nm). This difference could result either from the difference of charge transition rates or defect number and types (see the discussion of possible defect types in supporting information), which was natively formed during growth or transfer(*21*). A further determination would require the combination of SMLM with spectroscopic approaches(*33*) like photoluminescence, deep-level transient spectroscopy, or electron energy-loss spectroscopy. We foresee that such correlative techniques will provide further information on the electronic structures of individual defects.

In addition to chemical contrast, SMLM also offers quantitative information on the number of defects, which can be estimated based on their localizations (**Fig. 4A** and **Fig. 4B-E**). We also merged the localizations accounting for the localization error(*34*). The counting results from strict correction and total localizations offered both the bottom and upper limit of the defect density (**Fig. 4F-I**). Further improvement on this estimation can be made if the photophysics of defect emitters would be clearer. The obtained defect density (3268 $\mu m^{-2}$ -7246 $\mu m^{-2}$ for dense region **Fig. 4B** and 514 $\mu m^{-2}$ - 606 $\mu m^{-2}$ for sparse region **Fig. 4C**) of h-BN is comparable to previous results in ion transport(*35*) while quantitative SMLM avoids the averaging effect in ion transport measurements. We also used balanced SOFI (bSOFI) to obtain a density map of the defects as bSOFI (**SI Fig. 8**) was demonstrated to be more accurate than PALM for high-density emitters(*36*). This method provides a quantitative single molecule approach for estimating defect density in semiconductors on large areas.

A further interest to look at the quantum emitters is their spin properties, which can be optically read out when subject to electromagnetic fields, as shown in the measurements with nuclear quadrupole resonance spectroscopy(*37*). On the other hand, our method also provides a new approach to probe the properties of 2D materials since defects have been proven to perform a crucial role in determining their electrical transport and optical properties. To conclude, we have demonstrated the solid-state SMLM technique for imaging and counting quantum emitters in h-BN with the capability of distinguishing two closest emitters at a few nanometers. The achieved high localization precision may find future applications in semiconductor thin film characterization, quantum information processing, and biological imaging.

## Author Contributions

J.F conceived the idea and interpreted the results. J.F wrote the manuscript with inputs from A.R. J.F, H.D, and A.R designed experiments. J.F and H.D performed experiments. H.D processed all images. S.C and S.H performed materials growth. I.L. and P.L performed DFT calculations. All authors agreed with the final version of the manuscript. The authors declare no competing financial interest.

## Acknowledgment

This work was financially supported by Swiss National Science Foundation SNSF (200021 153653). We thank the Centre Interdisciplinaire de Microscopie Electronique (CIME) at the École Polytechnique fédérale de Lausanne (EPFL) for access to electron microscopes, Ke Liu, Davide Deiana, and Duncan T. L. Alexander for TEM imaging. Device fabrication was partially carried out at the EPFL Center for Micro/Nanotechnology (CMi). Special thanks to Martina Lihter for hBN transfer and Tomas Lukes for bSOFI image analysis which results in SI Fig. 8. The work performed in Cambridge was supported by the EPSRC Cambridge NanoDTC, EP/L015978/1. I. L. and P. L. were supported by the Unity Through Knowledge Fund, Contract No. 22/15 and H2020 CSA Twinning Project No. 692194, RBI-T-WINNING.

# Figures

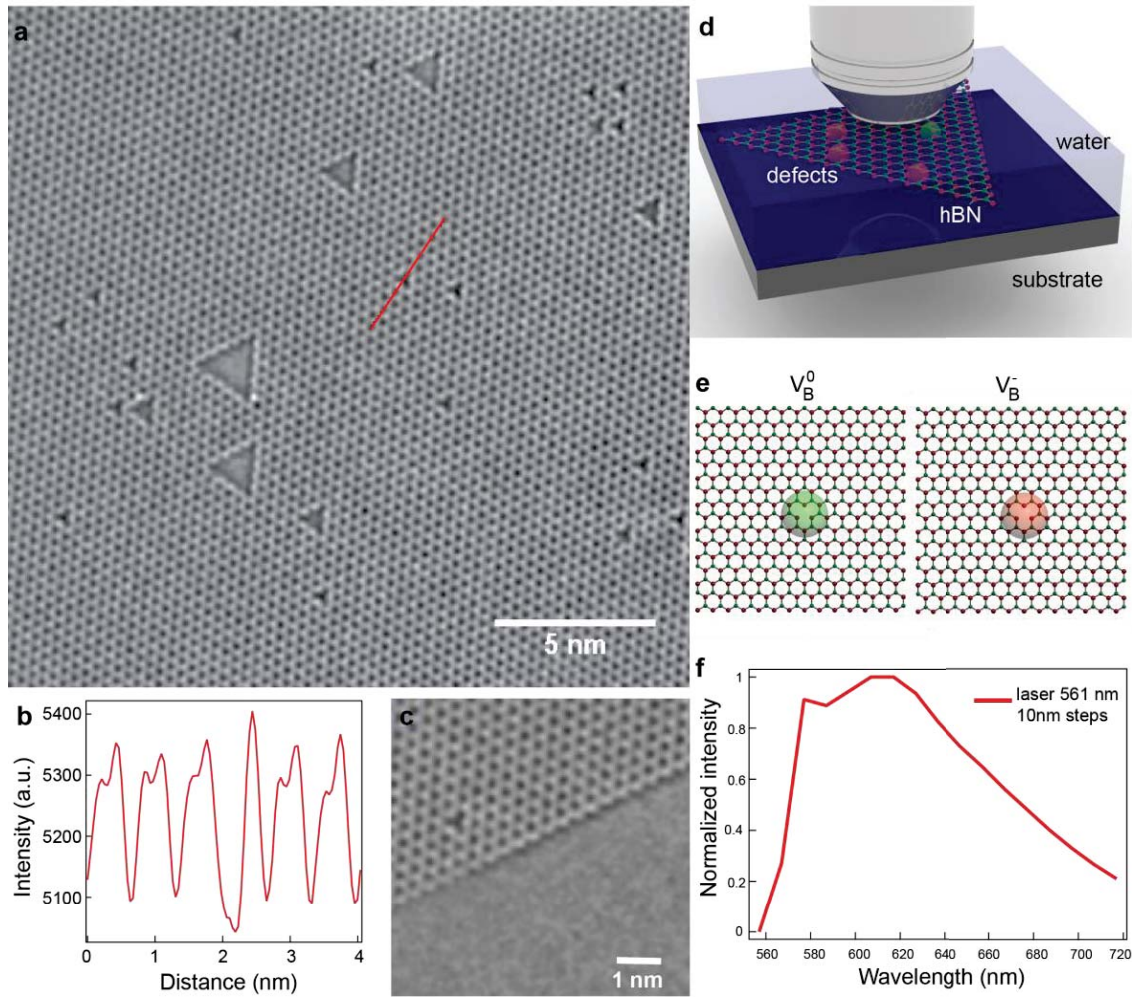

**Figure 1. Optically active defects in h-BN. A**, High-resolution TEM imaging of vacancy defects in h-BN. Dominant defects are found to be point defects. **B**, Image contrast of the line in **A** suggests the defect type to be boron monovacancy. **C**, High-resolution TEM imaging of h-BN edges. **D**, Schematics of *in-situ* characterization of defects in h-BN using SMLM (More details are shown in SI methods). **E**, Proposed structure of boron monovacancy with neutral and negative charge. **F**, Emission spectrum of h-BN defects.

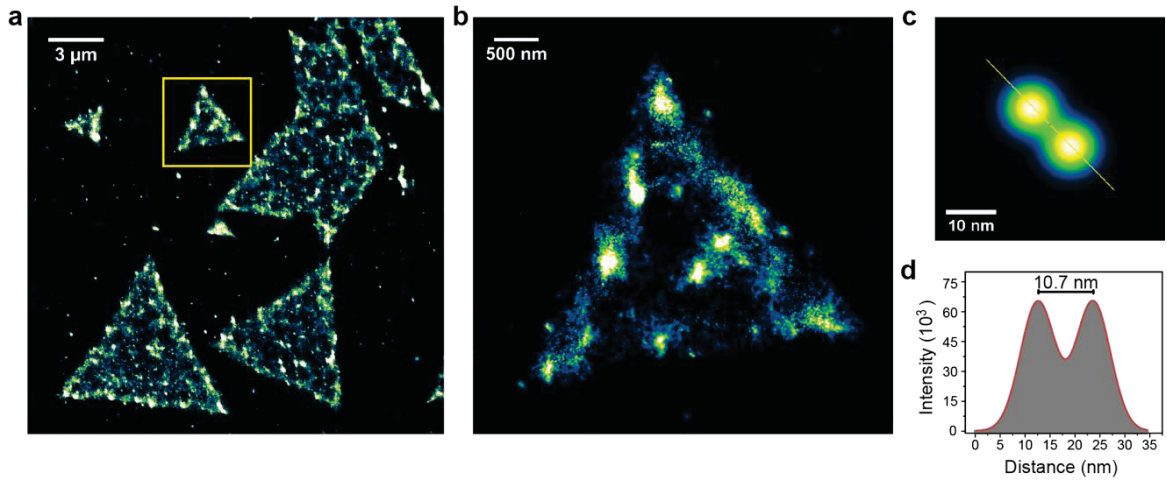

**Figure 2. Super-resolution imaging of optically active defects in h-BN. A**, Reconstructed SMLM image of defects in h-BN from an image sequence of 20, 000 frames. The imaging condition: DI water and 561 nm laser. Display pixel size: 5 nm. **B**, Selected region in **A**. Display pixel size: 1 nm. **C**, With best drift correction achieved with fiducial markers shown in **SI Fig. 3**, SMLM allows distinguishing two close emitters. **D**, Distance profile shown in **C** shows a resolving capability of 10.7 nm. The images were rendered as probability maps.

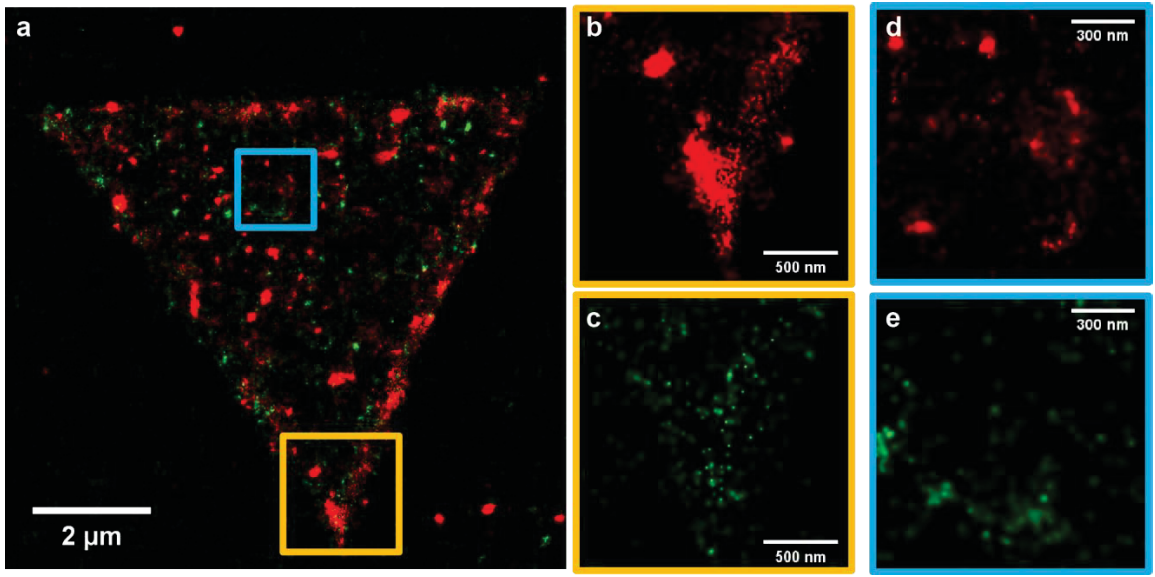

**Figure 3. Dual color imaging reveals chemical contrast. A**, Composited dual color super-resolution image of quantum emitters in h-BN from image sequence of 20, 000 frames. The imaging condition: DI water. Green channel: 488 nm laser. Red channel: 561 nm laser. **B, C, D, E**, Selected regions in **A**. The images were rendered as probability maps. Display pixel size: 5 nm.

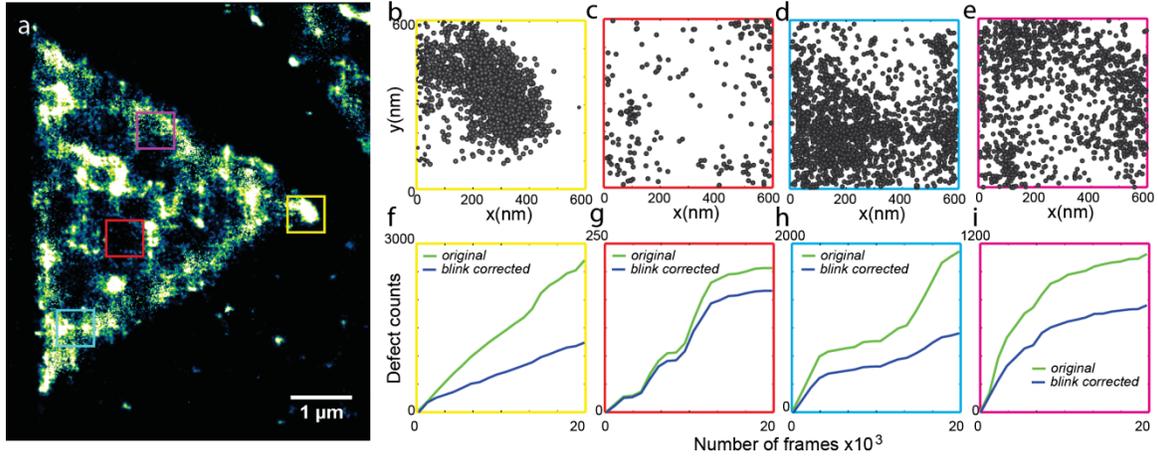

**Figure 4. Counting defects in h-BN using quantitative SMLM. A**, Super-resolution imaging and mapping of defect emitters in h-BN with color marked region of interest (ROI). **B, C, D, E,** localizations in color marked ROI in **A**. ROI size: 600 nm × 600 nm. **F, G, H, I,** Estimation of defect number in **B, C, D, E** as a function of image frame numbers using total number of localizations(original) and blink correction, respectively. The obtained density in **B**: 3268 μm$^{-2}$ (blink corrected) to 7246 μm$^{-2}$ (original). **C**: 514 μm$^{-2}$ (blink corrected) to 606 μm$^{-2}$ (original). **D**: 2849 μm$^{-2}$ (blink corrected) to 5208 μm$^{-2}$ (original). **E**: 2252 μm$^{-2}$ (blink corrected) to 3247 μm$^{-2}$ (original). Imaging condition: DI water and 561 nm laser.

# Imaging optically active defects with nanometer resolution

Jiandong Feng[1], Hendrik Deschout[1], Sabina Caneva[2], Stephan Hofmann[2], Ivor Lončarić[3], Predrag Lazić[3] and Aleksandra Radenovic[1]

[1]Laboratory of Nanoscale Biology, Institute of Bioengineering, School of Engineering, EPFL, 1015 Lausanne, Switzerland
[2]Department of Engineering, University of Cambridge, JJ Thomson Avenue, CB3 0FA Cambridge, United Kingdom
[3]Institut Ruđer Bošković, Bijenička 54, 10000 Zagreb, Croatia

*Correspondence should be addressed jiandong.feng@epfl.ch and aleksandra.radenovic@epfl.ch

# Supporting Information

**SI Text**

# SI Text

## 1. Sample preparation

### 1.1 CVD growth

The h-BN samples are grown under same conditions described elsewhere (*1*). Briefly, as-received Fe foil (100 μm thick, Alfa Aesar, 99.99% purity) is loaded in a customized CVD reactor (base pressure $1 \times 10^{-6}$ mbar). The foils are exposed to 4 mbar of $NH_3$ during heating to 900 °C. Subsequently, the $NH_3$ is removed and borazine $(HBNH)_3$ is introduced into the chamber through a leak valve from a liquid reservoir. After growth (45-90 s borazine exposure at $6 \times 10^{-4}$ mbar and 900 °C) the heater is switched off and the foils are cooled in a vacuum.

### 1.2 Transfer

The h-BN domains are transferred onto $SiN_x$ membranes using the electrochemical bubbling method (*2*). A polymer support layer is deposited onto the h-BN/Fe by spin coating poly(methyl methacrylate) (PMMA) at 5000 rpm for 40 s. The samples are placed in a NaOH bath (1M) and are contacted with a Pt wire that acts as the cathode, while another Pt wire (anode) is immersed in the electrolyte. During electrolysis, $H_2$ bubbles are generated at the h-BN/Fe interface, lifting the h-BN/PMMA film from the substrate. The film is subsequently rinsed in DI water and scooped with the target substrate, where it is allowed to dry. The PMMA is removed by immersing the sample in acetone overnight, followed by a rinse in IPA.

## 2. Imaging conditions

### 2.1 Microscope setup

Imaging was carried out on a custom-built microscope that was described previously (*3, 4*). Briefly, two laser sources were used for excitation: a 100mW 488 nm laser beam (Sapphire, Coherent) and a 100mW 561 nm laser beam (Excelsior, Spectra Physics). The laser beams were combined using a dichroic mirror (T495lpxr, Chroma) and sent through an acousto-optic tunable filter (AOTFnC-VIS-TN, AA Opto-Electronic). Both laser beams were focused into the back focal plane of an objective (UApo N ×100, Olympus) with a numerical aperture of 1.49 mounted on an inverted optical microscope (IX71, Olympus). Excitation and fluorescence light were separated by a filter cube containing a dichroic mirror (493/574 nm BrightLine, Semrock) and an emission filter (405/488/568 nm StopLine, Semrock). The fluorescence light was detected by an EMCCD camera (iXon DU-897, Andor) with a back-projected pixel size of 105 nm. An optical system (DV2, Photometrics) equipped with a dichroic mirror (T565lpxr, Chroma) was used to split the fluorescence light into a green and red color channel that were each sent to a separate half of the camera chip.

### 2.2 Imaging

Before imaging, 100 nm gold nanospheres (C-AU-0.100, Corpuscular) were added to the chip for lateral drift monitoring. The addition of fiducial markers was done by incubating the chip at least 15 minutes with a droplet of the gold nanospheres diluted up to 10 times in water. After removing the droplet, the chip was baked at 120 °C for at least 10 minutes and rinsed with water. The chip was placed upside down on a 25 mm diameter round cover slip (#1.5 Micro Coverglass, Electron Microscopy Sciences) that was cleaned with an oxygen plasma for 5 minutes. Imaging was performed in water at room temperature. Excitation in the red channel was done at 561 nm with ~20 to ~30 mW power (as measured in the back focal plane of the objective). Excitation in the green channel was done at 488 nm with ~10 mW power. The gain of the EMCCD camera was set at 100 and the exposure time to 50 ms. For each experiment, at least 20,000 camera frames were recorded. Dual-colour imaging was performed by first acquiring at least 10,000 camera frames in the red channel, and subsequently recording at least 10,000 camera frames in the green channel. Gold nanospheres were imaged to co-register the two color channels a posteriori. Axial drift correction was ensured by a nanometer positioning stage (Nano-Drive, Mad City Labs) driven by an optical feedback system (5)

## 2.3 Localization procedure

The camera frames were analyzed by a custom written algorithm (Matlab, The Mathworks) that was adapted from an algorithm that was described elsewhere (3, 4). First, a Gaussian filter and subsequently the Gaussian curvature filter (5) was applied to each frame separately. This yields an image of the uneven background consisting of triangular structures that correspond to fluorescence from the bulk hBN. This background image was subtracted from the original frame (after applying the same Gaussian filter), resulting in an image of the emitters on a uniform background. Only peaks with an intensity of at least 4 times the background were considered to be emitters. These peaks were fitted by maximum likelihood estimation of a 2D Gaussian point spread function (PSF) (6) and the localization precision was obtained from the Cramér-Rao lower bound (7). Drift was corrected in each frame by subtracting the average position of the gold nanospheres from the emitter positions in that frame. Co-registration of the two color channels was done using a second order polynomial transformation that was derived from the localizations of the gold nanospheres visible in both color channels. The SMLM images were generated as a probability map by plotting a 2D Gaussian PSF centered on each fitted position with a standard deviation equal to the corresponding localization precision. Only positions with a localization precision below 30 nm were plotted.

## 2.4 Counting

Determining the number of emitters by simply counting the localizations yields an overestimation in case there is "blinking," i.e. the same emitter reappears once or several times. We corrected this possible error by borrowing a method that was developed in the context of counting photo-activatable fluorescent proteins(8). The idea is that two different localizations are considered to originate from the same blinking event if they are close enough in space and time. Merging these localizations based on a suitable spatial and temporal threshold, therefore, results in a corrected number of localizations. To account for the localization precision, the spatial threshold was calculated from the Hellinger distance, which was taken equal to 0.9 (4). The temporal threshold was varied between multiples of the camera exposure time, i.e. the first five multiples (8), yielding five different corrected localization numbers. A semi-empirical model was fitted to these values as a function of the temporal threshold, resulting in a final corrected localization number (8). The model assumes that the emitters go from an off-state to an on-state, and subsequently they either reversibly go to a dark state or irreversibly to a photobleached state.

## 3. Defect electronic band diagrams

Electronic structure of vacancies in monolayer h-BN has previously been studied by Attaccalite et al.(*9*), Li et al.(*10*), and Tran et al.(*11*). Attaccalite et al. used both simple and rather inaccurate local density approximation (LDA) to density functional theory (DFT) and complex but usually very accurate the so-called GW/BSE theory. Li et al. and Tran et al. used generalized gradient approximation (GGA) to DFT, which in this case gives similar results to LDA. From these studies at LDA or GGA level, it is known that several type of defects have energy levels that can result in optically active transitions with energies of transition of around 2 eV that are observed in our and previous experiments. This is clearly shown in Fig. 4 of Ref. (*11*) for N monovacancy and B vacancy with the N atom shifted over into the empty site ($N_BV_N$), and also in Fig. 6 of Ref. (*10*) for B monovacancy, N monovacancy, divacancy and $N_BV_N$. However, it is well known that LDA and GGA have a tendency to underestimate band gaps of insulators, and therefore their predictions of optical properties of insulators are rather inaccurate(*12*). Additionally, DFT neglects excitonic effects. Band gap problem of LDA/GGA can be corrected by the so-called GW method, and excitonic effects can be included by solving the Bethe-Salpeter equation (BSE). Attaccalite et al. performed such calculation for B monovacancy, N monovacancy, and divacancy. Not taking into account excitonic effects, the lowest optically relevant transitions in GW for B monovacancy was obtained at 3.5 eV and for N monovacancy at 4 eV. When excitonic effects were included (GW/BSE), Attaccalite et al. predict that the lowest optical peak for B monovacancy should be at around 3.3 eV and for N monovacancy at 2.7 eV. The GW/BSE theory is probably the best method for solid state optical spectroscopy simulations, and therefore it is curious that it seems that this theory does not predict any defect related optical peak that would be in agreement with experiments. One could conclude that either neutral monovacancies are not responsible for peaks around 2 eV, or GW/BSE fails significantly for h-BN defects, or the theoretically studied structure of the system does not represent well the experimental conditions. To enlarge and deepen the search for defects observed in experiments we perform calculations for some of the observed defects in TEM images. Instead of using computationally very demanding GW/BSE, we use hybrid DFT functional that is known to largely correct the band gap problem of LDA/GGA(*12*).

Our TEM images show the uniform orientation of all defects. The most frequent defect is B monovacancy, but several larger vacancies are also visible. Due to these experimental facts, we have performed DFT calculations of B monovacancy and two larger vacancies with the same orientation, 3B+N and 6B+3N. Structures of these defects are shown in **Fig. S9**.

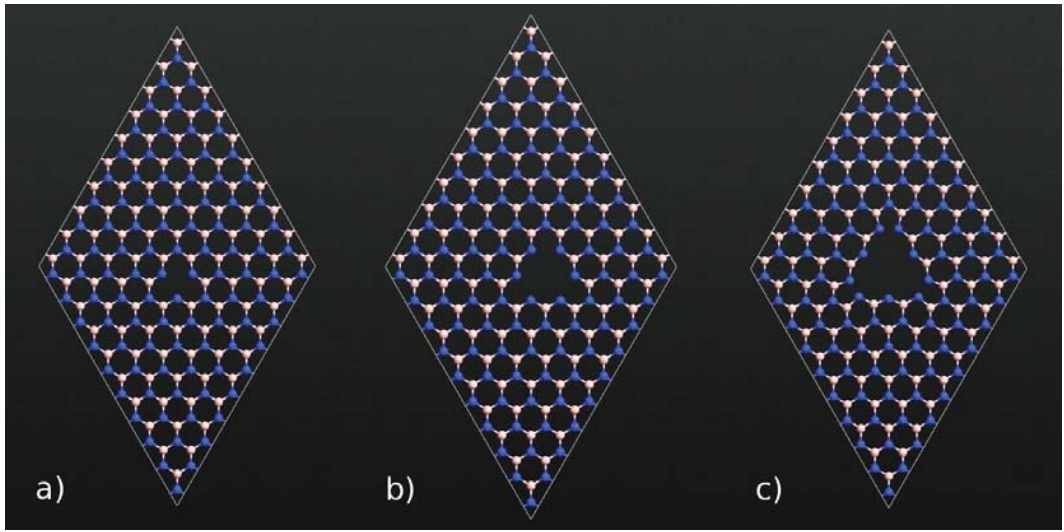

**Fig. S9.** PBE relaxed structure for a) B monovacancy, b) 3B+N vacancy, and c) 6B+3N vacancy.

All our calculations are performed using the plane-wave basis set VASP code(*13*). We use 10 × 10 BN unit cell to avoid interaction between defects and Γ point sampling of reciprocal space. For each defect, we first use commonly employed PBE exchange-correlation functional(*14*) to relax structures and then we perform hybrid DFT calculation using the so-called HSE06 functional(*15, 16*). As noted above, it is well-known that ordinary PBE functional usually underestimates band gaps and that it is prone to delocalization errors. As we want to simulate optical properties of defect levels in monolayer BN, both of these errors can severely limit the applicability of PBE to simulate this system. Hybrid functionals that include a portion of exact exchange usually give much more accurate band gaps and delocalization errors are smaller. Our DFT results are shown as band diagrams in **Fig. S10** for B monovacancy, **Fig. S11** for 3B+N vacancy, and **Fig. S12** for 6B+3N vacancy. Additionally, as we connect photo switching of strong emitters, observed in experiments, with charged defects, in **Fig. S10** we also show the calculation for charged B monovacancy. Our PBE results are consistent with previous studies(*9-11*). We observe large differences between PBE and HSE06 results for all three vacancies. Differences in band gaps are clearly visible both for bulk bands (bulk band gap is around 4.7 eV in PBE calculation and around 6.2 eV in HSE06 calculation) and defect levels. Moreover, there are some vacancy related levels inside the band gap that appear in the HSE06 calculations and not in the PBE calculations and *vice versa*. Due to the reasons stated above, we believe that HSE06 results should be much closer to reality and, therefore, in the following, we base our conclusions on them. To explain the experimental results, we should focus on defect levels that lead to transition energies of around 2 eV. Note that experimentally we measure spectra only in narrow energy range (1.65 eV to 2.75 eV) and, therefore, there might be additional, possibly stronger, optical peaks outside this range that are not seen in experiments.

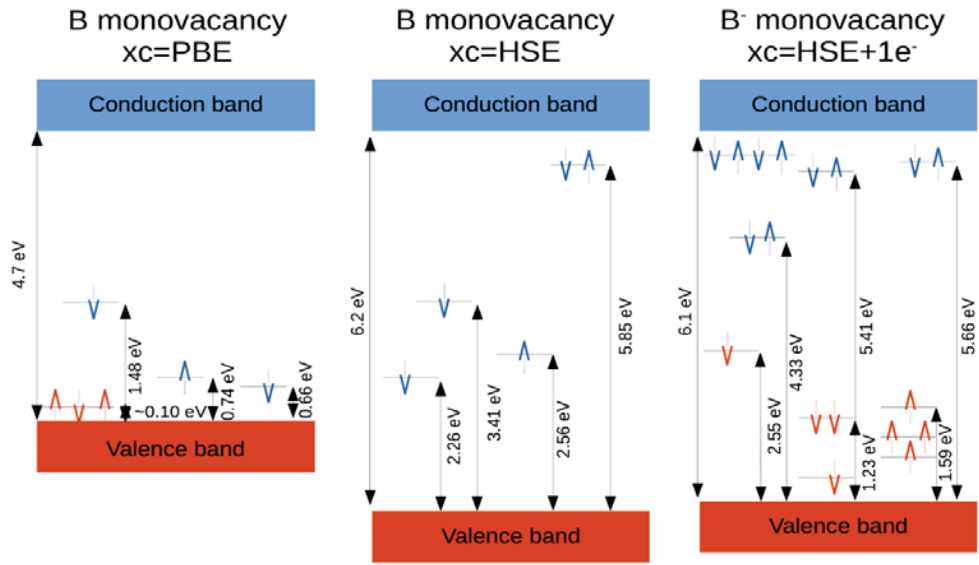

**Fig. S10.** Band diagrams of neutral (PBE and HSE06) and charged (HSE06) B monovacancy. Occupied states are in orange and unoccupied states are in blue.

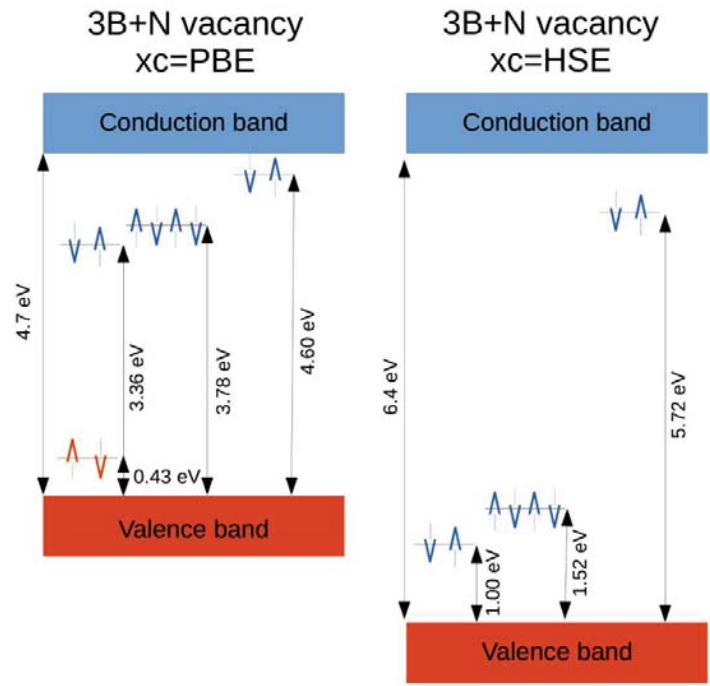

**Fig. S11.** Band diagrams of neutral 3B+N vacancy. Occupied states are in orange and unoccupied states are in blue.

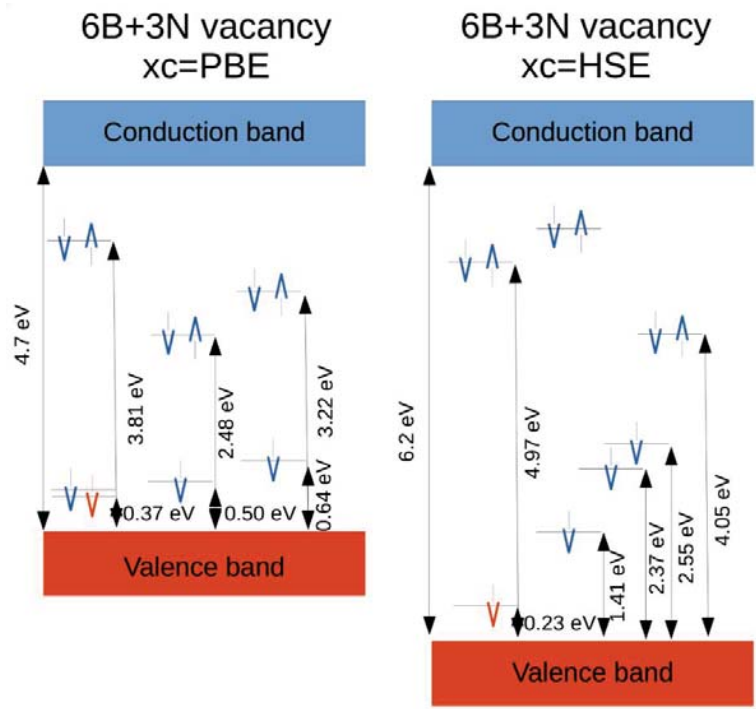

**Fig. S12**. Band diagrams of neutral 6B+3N vacancy. Occupied states are in orange and unoccupied states are in blue.

In addition to available transition energies, photoexcitation and fluorescence spectra also depend on transition strength between two states that is given by dipole transition matrix elements. Therefore, in **Fig. S13** we plot spectra that show both transition energies and transition strengths.

Results for each vacancy are listed below:

- **B monovacancy:** B monovacancy is the most common defect in our TEM images and therefore it is the first candidate for the homogeneous fluorescence. In agreement with previous studies, PBE results show three unoccupied defect levels at 0.66-1.48 eV above the valence band (left panel of **Fig. S10**). The highest one is responsible for optical transitions to states within valence band with the energy of around 2.3 eV as shown in the left panel of **Fig. S13**. Such peak in optical spectra would fit experiments. On the other hand, hybrid functional DFT results (middle panel in **Fig. S10**) also show three unoccupied defect levels within bulk band gap but with higher energies of 2.26 eV to 3.41 eV above the valence band. These levels have transition energies toward the top of the valence band that would also be in accordance with the experimental results, especially considering that excitonic effects would lower these transition energies. However, as can be seen in the right panel of **Fig. S13**, transition strengths for these transitions are very low and we obtain the same conclusions that were obtained in Ref. (*9*) with the GW/BSE theory. Namely, there are no significant optical transitions below 3.5 eV, far higher than observed in experiments. These results are puzzling and should be resolved by future measurements and calculations. From the one side, experimentally the most likely candidate for homogenuous emission is supported by PBE calculations that are typically rather inaccurate. From the other side, state of the art theoretical calculations like hybrid functional DFT or GW/BSE do not support it. Before concluding that B monovacancy is not responsible for homogeneous emission, further research is needed keeping in mind that even though GW and HSE are robust methods, it is possible that for some (still unknown) reasons they fail in predicting optical properties of this defect. Additionally, it should be explored whether defect-defect, defect-substrate, and defect-adsorbate interactions that are present in experimental conditions influence the electronic structure of the defect. From an experimental point of view, it would be helpful to have additional measurements in the full range of energies.

- **Charged B monovacancy:** Results for negatively charged B monovacancy (right panel in **Fig. S10**) show several additional occupied and unoccupied defect levels within the bulk band gap. The transition from the highest occupied to the lowest unoccupied defect level corresponds to the energy of 1.78 eV. Right panel of **Fig. S13** shows a clear peak at this transition energy, which demonstrates substantial transition strength. As this transition is due to the two defect levels within band gap it would be consistent with a single emitter. However, this transition energy is still significantly lower than in experiments, and the difference is larger than typical hybrid functional errors(*12*). It should be said that we did not additionally relax the charged defect compared to the neutral defect and therefore the forces on N atoms around the defect are rather large (~1 eV/Å). This could have an impact on the position of the defect levels. As can be seen in **Fig. S13**, there is an additional smaller absorption peak at around 2.8 eV that is also due to the transition between two defect levels. Since our results do not include excitonic effects that can easily shift transition energies down for 0.4 eV (see Ref. (*9*)), this peak would be consistent with experiments. All in all, it seems that charged B monovacancy could be responsible for ultra-bright emitters.

- **3B+N vacancy:** HSE band diagram of 3B+N vacancy show unoccupied defect level at 1.0 eV above valence band and double degenerate level at 1.5 eV above valence band. As can be seen in right panel of **Fig. S13** transition strengths to the top of the valence band are small and they become large only for transition energies larger than 2.5 eV. Therefore, hybrid functional DFT without considering excitonic effects (that would shift peak to lower energies) predict that 3B+N vacancy have a peak in optical spectra at around 2.8 eV. Due to non-zero dipole transition matrix elements at lower transition energies (toward the top of the valence band), fluorescence is expected to proceed also on lower transition energies. We conclude that 3B+N vacancy would be a good candidate for defect showing homogeneous fluorescence if its density would be high enough.

- **6B+3N vacancy:** For this vacancy, there is a transition with the energy of 1.2 eV which is clearly visible in the right panel of **Fig. S10**. In the spectra there is also a peak around 1.5 eV which corresponds to transitions from the defect level at 1.41 eV above valence band to the levels inside the valence band. Some small peaks also exist at transition energies in range 2-2.5 eV that could also be relevant considering the experiments, but the density of this type of vacancy is probably too low.

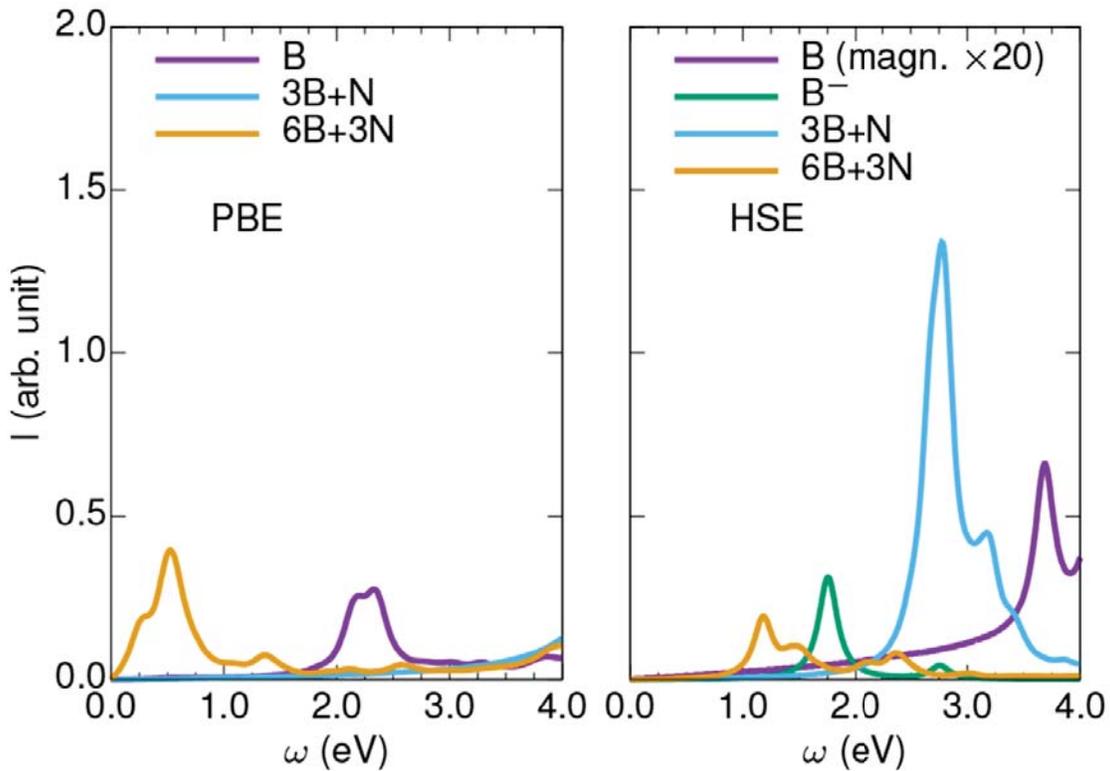

**Fig. S13.** Simulated optical spectra in independent particle approximation calculated with PBE (left panel) and HSE06 (right panel) functional.

## 5. SI Figure Captions

**Fig. S1.** Generated diffraction-limited image by summing all 20,000 frames, corresponding to the SMLM image shown in **Fig. 2a**.

**Fig. S2.** FRC resolution analysis. **A.** Generated SMLM image using even number frames. **B.** Generated SMLM image using odd number frames. Total number of frames: 20,000. Image conditions: DI water, 561 nm laser. **C.** FRC curve indicates the decay of the correlation with spatial frequency. The image resolution (46 nm) is obtained using the inverse of the spatial frequency where FRC curve drops below the threshold 1/7.

**Fig. S3.** Drift correction using fiducial markers. **A.** Corrected position of 30 fiducial markers (100 nm gold nanoparticles- from which 27 fiducial markers are used for drift correction). **B.** Drift trajectories of all 27 fiducial markers over 20,000 frames. **C** and **D.** x,y stage drift as a function of frame numbers.

**Fig. S4.** Log-scale histogram of photon counts for the results shown in **Fig. 2A**. Insets show zoomed region of high photon counts.

**Fig. S5.** SMLM imaging of defects in h-BN under basic and acidic conditions. **A.** SMLM image acquired under basic environment (pH 11, water).. **B** and **C**, Selected ROI in **A** and corresponding counting results **D** and **E**, respectively. The average defect density is from 1018 $\mu m^{-2}$ (blink corrected) to 1648 $\mu m^{-2}$ (total localizations). **F.** SMLM image acquired under acidic environment (pH 3, water). **G** and **H.** Selected ROI in **F** and corresponding counting results **I** and **J**, respectively. The average defect density is from 198 $\mu m^{-2}$ (blink corrected) to 307 $\mu m^{-2}$ (total localizations).

**Fig. S6.** SMLM imaging of defects in h-BN after surface deposition. **A.** SMLM image of h-BN sample where defects are sealed by depositing 2 nm $Al_2O_3$ with atomic layer deposition. **B** and **C,** Selected ROI in **A** and corresponding counting results **D** and **E**, respectively. The average defect density is from 201 $\mu m^{-2}$ (blink corrected) to 253 $\mu m^{-2}$ (total localizations).

**Fig. S7.** SMLM imaging of defects in h-BN under isotopic solvent conditions. **A.** SMLM image of defects in h-BN in $D_2O$ solvent. **B** and **C**, Selected ROI in **A** and corresponding counting results **D** and **E**, respectively. The average defect density is from 2186 $\mu m^{-2}$ (blink corrected) to 4132 $\mu m^{-2}$ (total localizations).

**Fig. S8.** Balanced super-resolution optical fluctuation (bSOFI) images obtained from the same raw image sequences as in **Fig 2a.**

**Fig. S1**

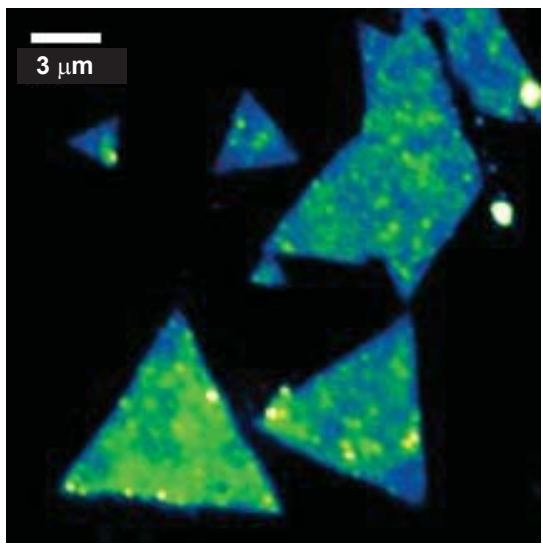

**Fig. S2**

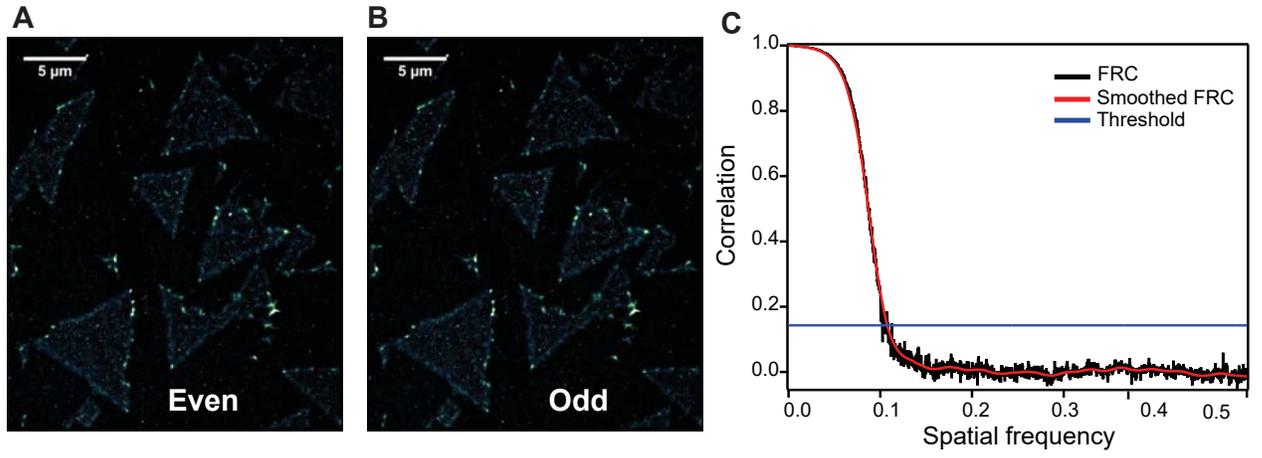

**Fig. S3**

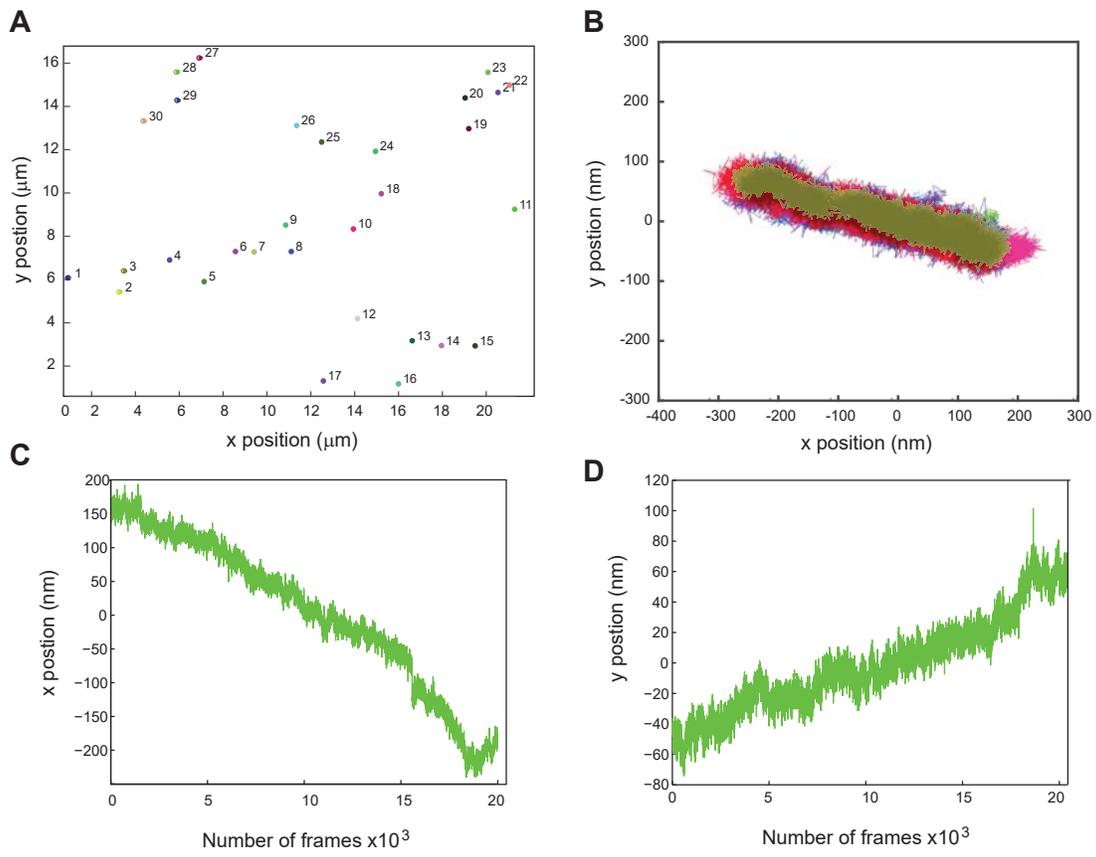

**Fig. S4**

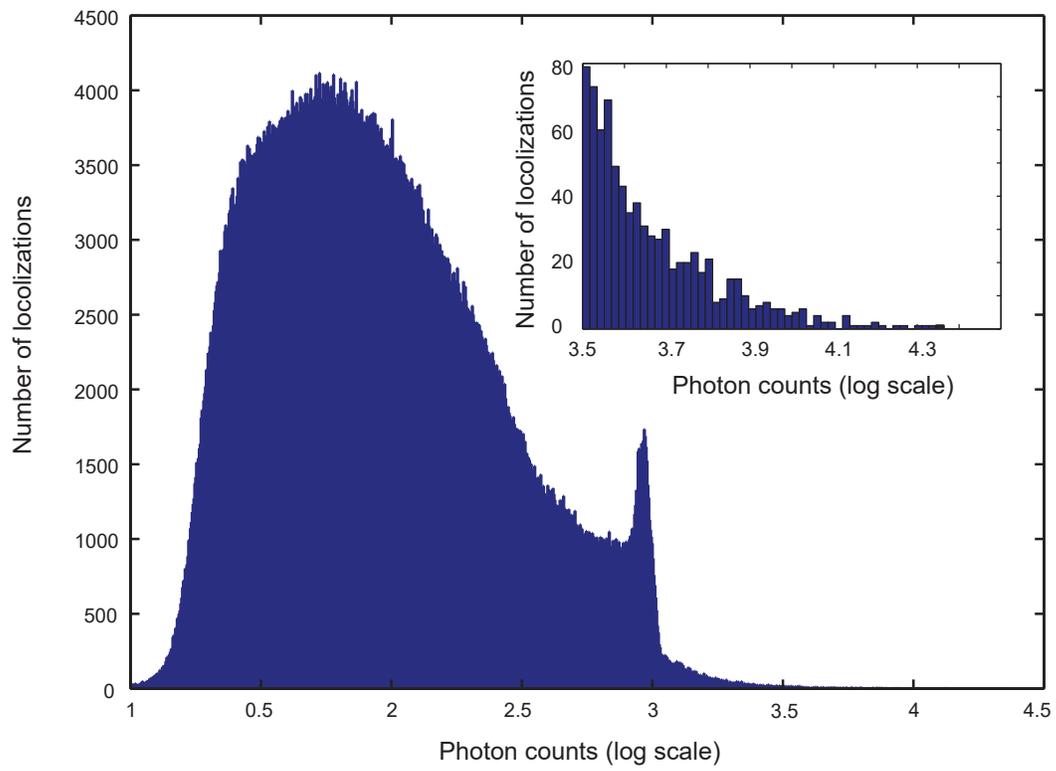

**Fig. S5**

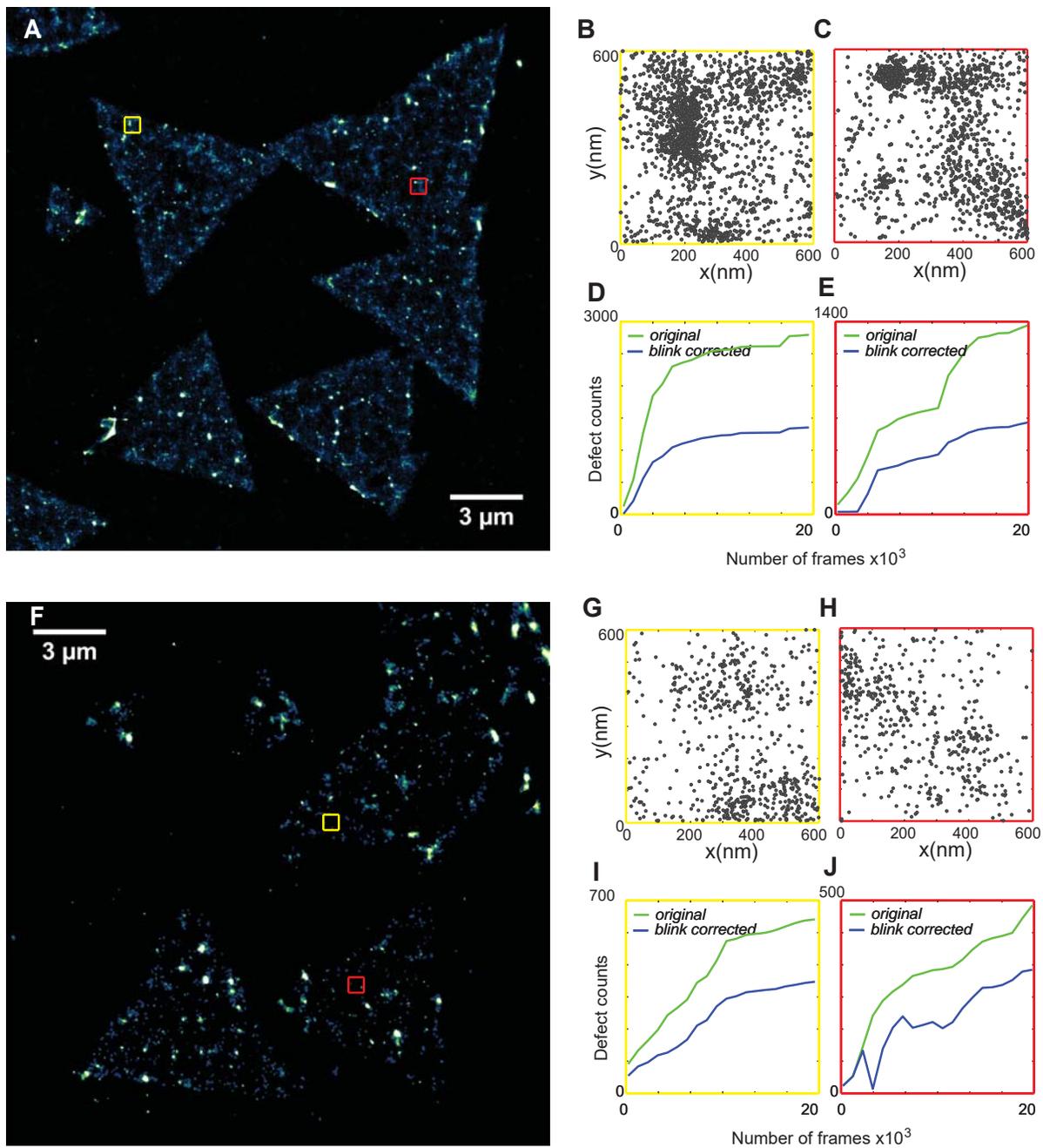

**Fig. S6**

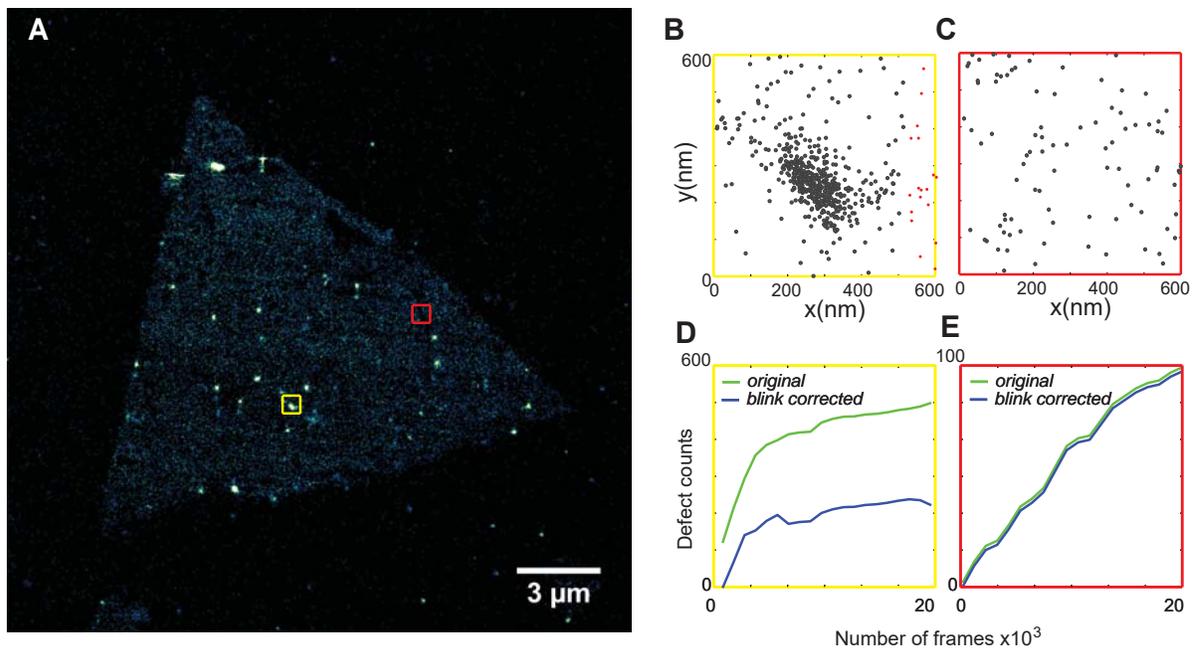

Fig. S7

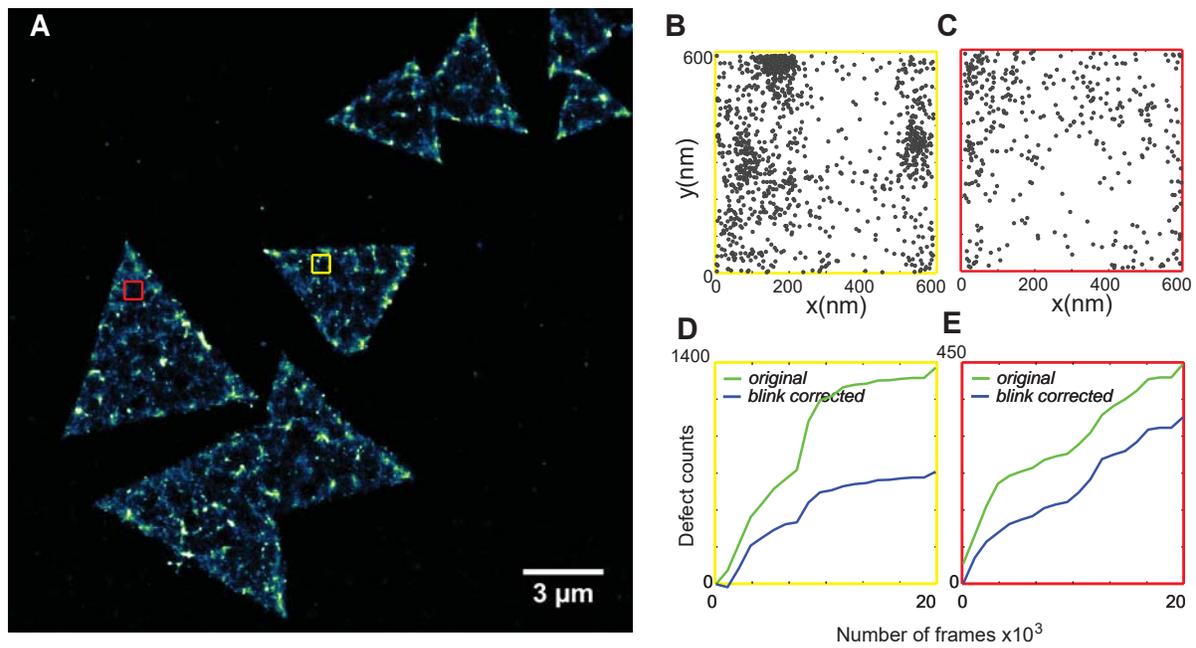

**Fig. S8**

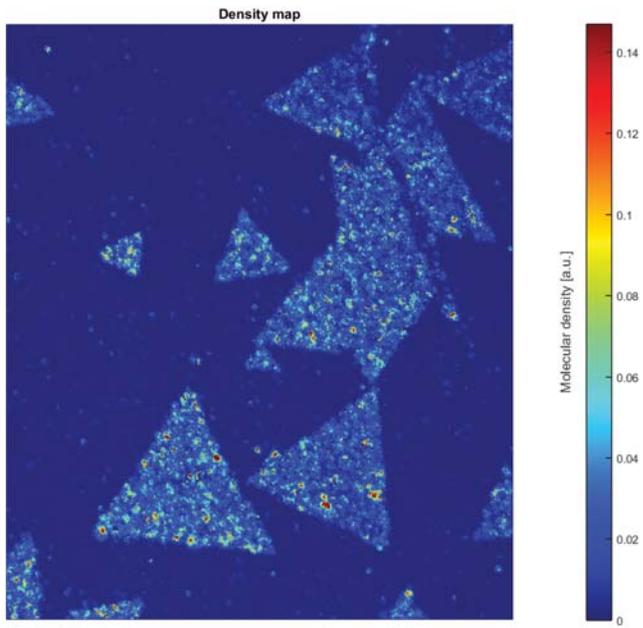